\documentclass[prb,twocolumn,showpacs,preprintnumbers,amsmath,amssymb]{revtex4-1}
\usepackage{graphicx}% Include figure files
\usepackage{epstopdf}
\usepackage{color}
\usepackage{bbold}

\begin{document}
\title{Inhomogeneous multi-carrier superconductivity at LaXO$_3$/SrTiO$_3$ (X=Al or Ti) oxide interfaces} 
\author{S. Caprara$^{1,2}$, D. Bucheli$^1$, N. Scopigno$^1$, N. Bergeal$^3$, J. Biscaras$^4$, S. Hurand$^3$,
J. Lesueur$^3$, and M. Grilli$^{1,2}$}
\affiliation{
$^1$Dipartimento di Fisica Universit\`a di Roma Sapienza, piazzale Aldo Moro 5, 
I-00185 Roma, Italy\\
$^2$ISC-CNR and Consorzio Nazionale Interuniversitario per le Scienze Fisiche della 
Materia, Unit\`a di Roma Sapienza, Italy\\
$^3$LPEM-UMR8213/CNRS-ESPCI Paris Tech-UPMC, 10 rue Vauquelin, 75005 Paris, France\\
$^4$IMPMC-UMR7590/CNRS-UPMC Case 115, 4 Place Jussieu, 75252 Paris, France}
%\author{} 
%\affiliation{}   
\begin{abstract} 
Several experiments reveal the inhomogeneous character of the superconducting state that occurs when the 
carrier density of the two-dimensional electron gas formed at the LaXO$_3$/SrTiO$_3$ (X=Al or Ti) interface 
is tuned above a threshold value by means of gating. Re-analyzing previous measurements, that highlight the 
presence of two kinds of carriers, with low and high mobility, we shall provide a description of multi-carrier 
magneto-transport in an inhomogeneous two-dimensional electron gas, gaining insight into the properties of 
the physics of the systems under investigation. We shall then show that the measured resistance, 
superfluid density, and tunneling spectra result from the percolative connection of superconducting ``puddles'' 
with randomly distributed critical temperatures, embedded in a weakly localizing metallic matrix. We shall 
also show that this scenario is consistent with the characteristics of the superconductor-to-metal transition 
driven by a magnetic field. A multi-carrier description of the superconducting state, within a weak-coupling 
BCS-like model, will be finally discussed.
\end{abstract}   
\date{\today} 
\maketitle 
   
\section{Introduction}
\label{sec1}
\begin{figure}
\includegraphics[scale=0.4]{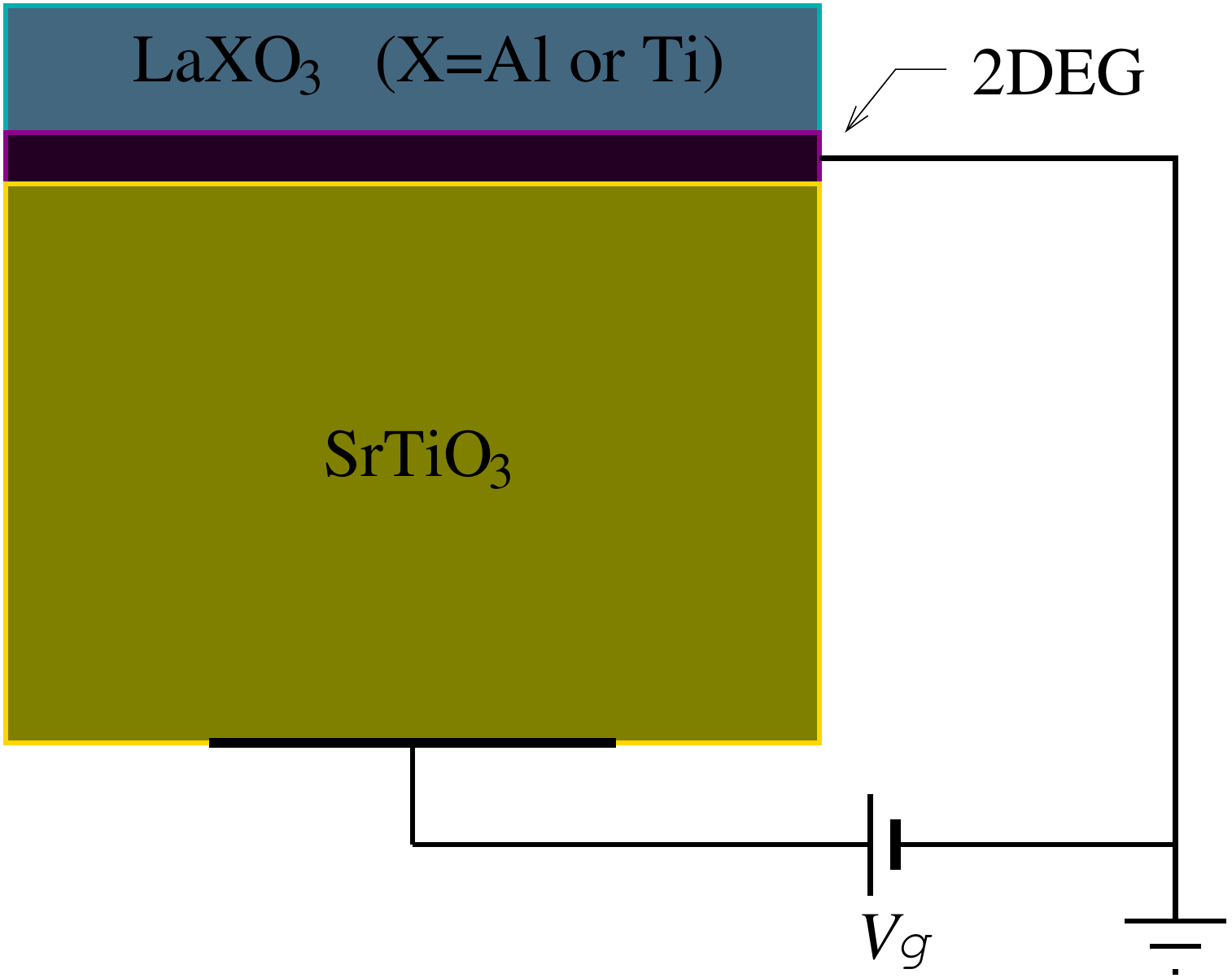}
\caption{Scheme of an oxide interface (not in scale), in the back gating configuration. The LAO (or LTO) topmost 
slab has a thickness of few nm, whereas the STO slab is $\approx 0.5$\,mm thick. The back gate voltage $V_g$ is 
employed to tune the carrier density of the 2DEG. The top gating configuration is also possible.}
\label{fig1}
\end{figure}

After a two-dimensional electron gas (2DEG) was detected at the interface between two insulating 
oxides,\cite{reyren,triscone,espci1,espci2} an increasingly intense theoretical and experimental
investigation has been devoted to these systems. The properties of this 2DEG are intriguing within several 
respects: it can be made superconducting when the carrier density is tuned by means of gate voltage (see 
Fig.\,\ref{fig1}), both in LaAlO$_3$/SrTiO$_3$ (henceforth, LAO/STO)\cite{reyren,triscone} and 
LaTiO$_3$/SrTiO$_3$ (henceforth, LTO/STO)\cite{espci1,espci2} interfaces, thus opening the way to voltage-driven 
superconducting devices; it exhibits magnetic properties;\cite{ariando,luli,bert,metha1,metha2,bert2012} it 
displays a strong and tunable\cite{cavigliaprl,CPG} Rashba spin-orbit coupling;\cite{rsoc} it is extremely 
two-dimensional, having a lateral extension $\approx 5$\,nm, thereby enhancing the effects of disorder due 
to extrinsic and/or intrinsic\cite{CPG} sources. 

Magneto-transport experiments reveal the presence of two kinds of carriers in LTO/STO, with high and low mobility, 
and superconductivity seems definitely to develop as soon as high-mobility carriers appear,\cite{espci2,bell} when 
the carrier density is tuned above a threshold value, by means of gate voltage $V_g$. When the temperature $T$ is 
lowered, the electrical resistance is reduced, and signatures of a superconducting fraction are seen well above 
the temperature at which the global zero resistance state is reached (if ever). The superconducting fraction 
decreases with decreasing $V_g$, although a superconducting fraction survives at values of $V_g$ such that the 
resistance stays finite down to lowest measured temperatures. When $V_g$ is further reduced, the superconducting 
fraction eventually disappears, and the 2DEG stays metallic at all temperatures, seemingly undergoing weak 
localization at low $T$. At yet smaller carrier densities, the system behaves as an insulator. The width of the 
superconducting transition is anomalously large and cannot be accounted for by reasonable superconducting 
fluctuations.\cite{CGBC} This phenomenology suggests instead that an inhomogeneous 2DEG is formed at these 
oxide interfaces, consisting of superconducting ``puddles'' embedded in a (weakly localizing) metallic background, 
opening the way to a percolative superconducting transition.\cite{BCCG} Inhomogeneities are revealed in various 
magnetic experiments,\cite{ariando,luli,bert,bert2012} in tunneling spectra,\cite{ristic} and in piezoforce 
microscopy measurements.\cite{feng_bi} Seemingly, inhomogeneities at the nanometric scale coexist with larger 
(e.g., micrometric) scale inhomogeneities, revealed by the occurrence of striped textures in the current 
distribution\cite{Moller} and in the surface potential.\cite{ilani2}

Various aspects of the phenomenology of oxide interfaces (henceforth referred to as LXO/STO interfaces, when 
referring to both LAO/STO and LTO/STO) have been separately discussed 
before.\cite{espci2,CPG,CGBC,BCCG,espci3,rapid} Here, we first provide new compelling evidence of the 
inhomogeneous character of the 2DEG, extending previous multi-carrier analyses of magneto-transport measurements 
to deal with inhomogeneous systems. We then put together the various pieces of the jigsaw into an overall 
coherent theoretical framework.

The plan of the present piece of work is the following. In Sec.\,\ref{sec2}, we propose a model for
multi-carrier magneto-transport in inhomogeneous systems, and show that previous analyses of the
magnetoresistance and Hall resistance measurements in terms of two different species of 
carriers\cite{espci2} is fully consistent with the inhomogeneous character of the 2DEG at LAO/STO and
LTO/STO interfaces. In Secs.\,\ref{sec3}, \ref{sec4}, and \ref{sec5}, we revisit some of our 
previous results. Assuming inhomogeneity as an empirical evidence, we show that 
resistance measurements\cite{espci2} and the topographic mapping of the superfluid density\cite{bert2012} 
can be accounted for within a percolative scheme. In Sec.\,\ref{sec6}, we discuss some preliminary
aspects of a theory for metal-insulator-superconductor tunneling in inhomogeneous superconductors that 
is apt to reproduce the measured tunneling spectra.\cite{tun} In Sec.\,\ref{sec7}, we provide further 
evidence for inhomogeneous superconductivity at oxide interfaces, coming from the peculiar multiple quantum critical 
scaling, observed when superconductivity is suppressed by means of a magnetic field, revisiting the
results of Ref. [\onlinecite{espci3}].
In Sec.\,\ref{sec8}, relying on the results of Ref. [\onlinecite{rapid}], we show that the properties of the 
superconducting puddles (e.g., their fraction, and critical temperatures) can be extracted from experiments 
and used to model intra-puddle multi-carrier superconductivity, gaining insight about the pairing mechanism. 
Although some features of the diamagnetic response are seemingly related to strong superconducting 
coupling,\cite{bert2012} we show that inhomogeneities and multi-carrier superconductivity fully account for 
the behavior of these systems within a standard weak coupling BCS scheme. Concluding remarks are found in 
Sec.\,\ref{sec9}.

\section{Multi-carrier magneto-transport in inhomogeneous systems}
\label{sec2}
%%%%%%%%%%%%%%%%%%%%%%%%%%%%%%%%%%%%%%%%%%%%%%%%%%%%%%%%%%%%%%%%%%%%%%%%
\begin{figure}
\vspace{0 truecm}
\includegraphics[angle=0,scale=0.3]{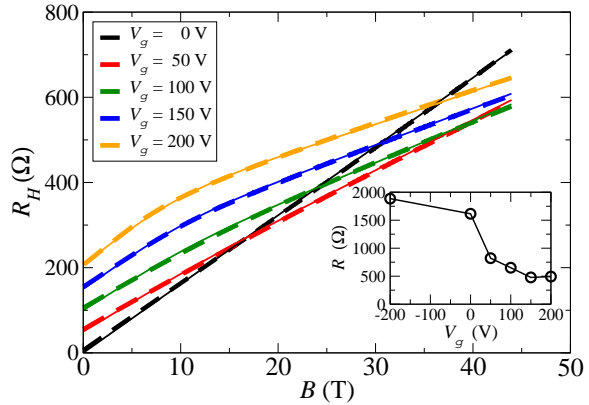}
\caption{Hall resistance as a function of magnetic field for different gate voltages $V_g$, measured at $T=4.2$\,K 
in a LTO/STO 15-unit-cell thick sample (sample A of Ref. [\onlinecite{espci2}]).
Solid lines  correspond to experimental data, taken from Ref. [\onlinecite{espci2}],
while the dashed lines are obtained here, fitting the data with Eq. (\ref{hallconst}). 
For clarity, the curves have been shifted upwards by 50\,$\Omega$ in increasing voltage order.
Inset: Sheet resistance $R$, obtained via Eq.\,(\ref{sigmaxx}), as a function of the gate voltage.}
\label{fig2}
%       fig. 4:
\end{figure}
%%%%%%%%%%%%%%%%%%%%%%%%%%%%%%%%%%%%%%%%%%%%%%%%%%%%%%%%%%%%%%%%%%%%%%%%

The detection of two species of carriers, with high and low mobility, by means of magneto-transport measurements
in LTO/STO,\cite{espci2} is not necessarily a direct evidence of inhomogeneity, since the two species could coexist 
in a homogeneous manner. However, to account for the phenomenology of the superconducting state of the 2DEG
at LXO/STO interface, we proposed that the 2DEG is inhomogeneous, with higher density regions (the
superconducting puddles) and lower density regions (the metallic background).\cite{espci3,rapid} Assigning a
band structure to the system, one is then led to assume that, when the density is large enough to fill the
bands occupied by the high-mobility carriers, these should locally coexist with low-mobility carriers occupying
the lower-lying levels (see also Sec.\,\ref{sec8}). A picture then emerges in which the low-mobility carriers
alone are present in the metallic background, whereas they coexist with the high-mobility carriers in the
superconducting puddles. To improve our description of LXO/STO interfaces, we  must then rely on a theory for 
multi-carrier magneto-transport that is apt to deal with an inhomogeneous system. One such theory has been developed 
in the form of an Effective Medium Theory (EMT) for the Hall conductance of a binary medium 
resulting from the mixture of two phases, based on rotation transformations (see Ref.\,[\onlinecite{arkhincheev}] for 
a detailed description of this method). 

In our description of the LXO/STO interfaces, one (less dense) phase hosts one species of carriers, 
with low mobility, and the other (denser) phase hosts two species of carriers, with low and high mobility.
Following Ref.\,[\onlinecite{arkhincheev}], we first define the conductivity tensor for each 
of the two coexisting phases, in the presence of a magnetic field of amplitude $B$. The diagonal elements of
the conductivity tensor are
\begin{eqnarray*}
\sigma^{(1)}_{xx}(B)&=&\frac{\sigma_1}{1+[\beta_1(B)]^2}, \nonumber \\
\sigma^{(2)}_{xx}(B)&=&\frac{\sigma_1}{1+[\beta_1(B)]^2}+\frac{\sigma_2}{1+[\beta_2(B)]^2}, \nonumber
\end{eqnarray*}
and the off-diagonal elements are
\begin{eqnarray*}
\sigma^{(1)}_{xy}(B)&=&\frac{\sigma_1 \beta_1(B)}{1+[\beta_1(B)]^2}, \nonumber \\
\sigma^{(2)}_{xy}(B)&=&\frac{\sigma_1 \beta_1(B)}{1+[\beta_1(B)]^2}+\frac{\sigma_2\beta_2(B)}{1+[\beta_2(B)]^2}. 
\nonumber 
\end{eqnarray*}

Hereafter, the superscripts $(1)$ and $(2)$ label the two phases, the subscripts $1$ and $2$ label the two 
species of carriers (with low and high mobility, respectively), and we adopt the notations
$\beta _{i}(B)\equiv\mu_i B/c$ and $\sigma_i\equiv e n_i\mu_i$, where $n_i$ is the carrier density and $\mu_i$ 
is the mobility of the $i$-th species of carriers. We point out that the peculiar aspect of our present
description is that both species of carriers contribute in parallel to the conductivity tensor in 
the phase (2), which we identify with the higher density superconducting puddles. We indicate with
$w\equiv \frac{1}{2}-\epsilon$ the fraction of the system occupied by the superconducting puddles, 
$\epsilon$ being the deviation from the percolation threshold, that changes sign when the minority phase 
($w<\frac{1}{2}$, $\epsilon >0$) percolates and becomes the majority phase ($w>\frac{1}{2}$, $\epsilon <0$), 
the threshold being $w=\frac{1}{2}$ in two-dimensional systems.
%%%%%%%%%%%%%%%%%%%%%%%%%%%%%%%%%%%%%%%%%%%%%%%%%%%%%%%%%%%%%%%%%%%%%%%%
\begin{figure}
\vspace{0 truecm}
\includegraphics[angle=0,scale=0.3]{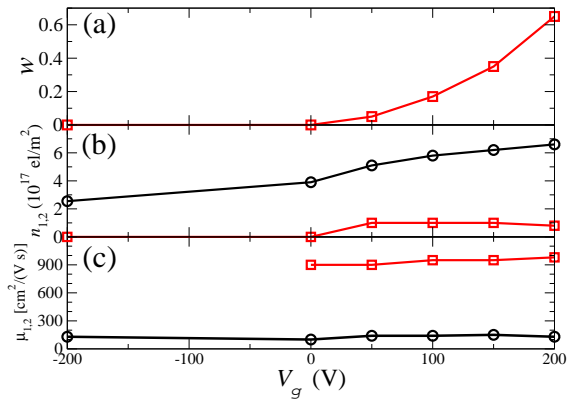}
\caption{(a) Fraction $w$ of the  high-mobility metallic phase [phase (2), see text];
(b) Densities of the less ($n_1$, empty circles) and more ($n_2$, empty squares) mobile carriers
extracted from the fits of the Hall resistance, Eq.\,(\ref{hallconst}), as a function of gate voltage measured 
at $T=4.2$\,K. 
(c) Mobilities $\mu_1$ (empty circles) and $\mu_2$ (empty squares) of the majority (low-mobility) and minority 
(high-mobility) carriers, respectively, extracted from 
Eq.\,(\ref{hallconst}). Type 1 (low-mobility) carriers only are present in the phase (1), while
both type 1 and type 2 (high-mobility) carriers are present in the phase (2).}
\label{fig3}
%       fig. 4:
\end{figure}
%%%%%%%%%%%%%%%%%%%%%%%%%%%%%%%%%%%%%%%%%%%%%%%%%%%%%%%%%%%%%%%%%%%%%%%%

Within EMT, the diagonal element of the conductivity tensor of the inhomogeneous (two-phase) system
can be found, that correctly reproduces the limiting case of a pure phase, and takes the form\cite{arkhincheev} 

\begin{eqnarray}
&&\sigma^{EMT}_{xx}(B,\epsilon)=\epsilon \left[\sigma^{(2)}_{xx}(B)-\sigma^{(1)}_{xx}(B)\right] \nonumber \\
&+& \sqrt{\epsilon^2\left[\sigma^{(1)}_{xx}(B)-\sigma^{(2)}_{xx}(B)\right]^2+\sigma^{(1)}_{xx}(B)\sigma^{(2)}_{xx}(B)}.
~~~\label{sigmaxx}
\end{eqnarray}

Then, exploiting duality relations that connect the various elements of the conductivity tensor when the 
minority and majority phases are interchanged (see Ref.\,\onlinecite{arkhincheev}), one obtains the EMT expression
for the off-diagonal element of the conductivity tensor of the inhomogeneous (two-phase) system
\begin{widetext}
\begin{eqnarray*}
&&\sigma^{EMT}_{xy}(B,\epsilon)=\left\{
\frac{\left[ \sigma^{EMT}_{xx}(B,\epsilon)+\sigma^{EMT}_{xx}(B,-\epsilon)  \right]
\left[ \sigma^{(1)}_{xx}(B)\sigma^{(2)}_{xy}(B) 
+\sigma^{(1)}_{xy}(B)\sigma^{(2)}_{xx}(B) \right]}{\sigma^{(1)}_{xx}(B)+\sigma^{(2)}_{xx}(B)}
\right. \nonumber \\
&&\left. -\frac{\left[ \sigma^{EMT}_{xx}(B,\epsilon)-\sigma^{EMT}_{xx}(B,-\epsilon)  \right]
\left[ \sigma^{(1)}_{xx}(B)\sigma^{(2)}_{xy}(B) 
-\sigma^{(1)}_{xy}(B)\sigma^{(2)}_{xx}(B) \right]}{\sigma^{(1)}_{xx}(B)-\sigma^{(2)}_{xx}(B)}\right\} 
\frac{1}{2\sigma^{EMT}_{xx}(B,-\epsilon)}.\nonumber
\end{eqnarray*}
\end{widetext}
From the above relations, the expression for the Hall resistance of the inhomogeneous system can 
finally be derived,
\begin{equation}
R_H^{EMT}(B,\epsilon)=\frac{\sigma^{EMT}_{xy}(B,\epsilon)}{[\sigma^{EMT}_{xy}(B,\epsilon)]^2
+[\sigma^{EMT}_{xx}(B,\epsilon)]^2},
\label{hallconst}
\end{equation}
which we used to accurately fit the experimental Hall resistivity data of LTO/STO,\cite{espci2} under strong 
magnetic field and at different gate voltages (see Fig.\,\ref{fig2}). This procedure allows to extract the values 
of the mobilities $\mu_i$ and of the densities $n_i$ of the two species of carriers, as well as the fraction of the 
system occupied by the superconducting puddles, $w$. These values are reported in Fig.\,\ref{fig3}, while the 
inset of Fig.\,\ref{fig2} displays the evolution of the normal-state sheet resistance, extracted via
Eq.\,(\ref{sigmaxx}) as $R=[\sigma^{EMT}_{xx}(B=0,\epsilon)]^{-1}$, with changing $V_g$. 
\begin{figure*}
\begin{center}
\includegraphics[scale=0.375]{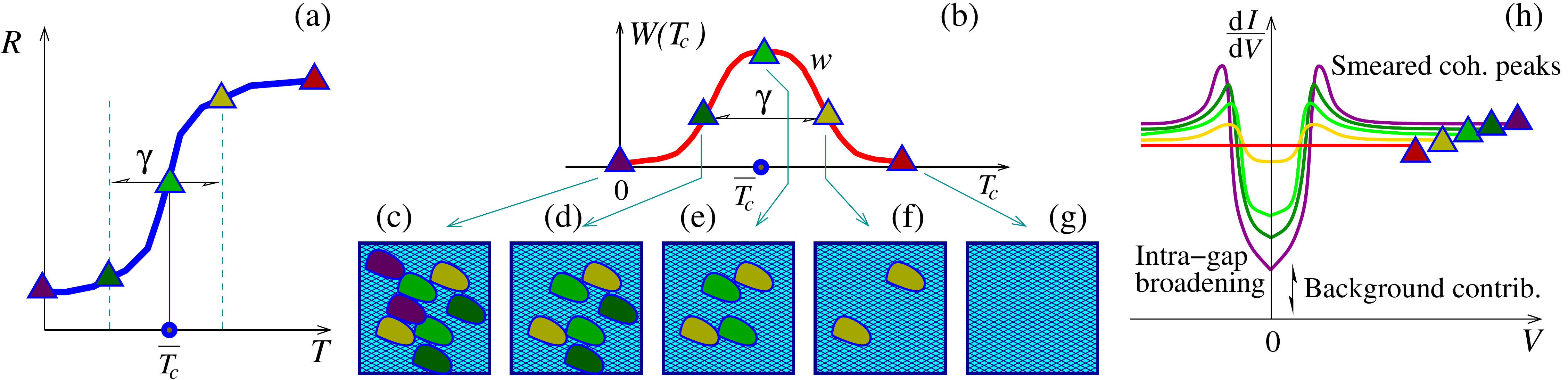}
\end{center}
\caption{(a) Sketch of the behavior of the resistance as a function of temperature in a case when 
the superconducting fraction does not percolate. (b) Distribution of critical temperatures $W(T_c)$ in the
superconducting puddles, that occupy a fraction $w$ of the sample (the remaining $1-w$ fraction will never 
become superconducting). 
(c$-$g) When the temperature is reduced [from (g) to (c)] superconducting puddles appear in the system as soon 
as the local critical 
temperature exceeds $T$. However, if the puddles do not percolate down to $T=0$, the global zero resistance state 
is never reached. (h) Sketch of the tunneling spectra measured in LAO/STO, as a function of bias voltage $V$: 
At temperatures well above the temperature
at which the zero resistance state is reached (if ever), a suppression is observed in the tunneling spectra around 
zero bias, accompanied by more or less pronounced coherence peaks above the gap. The coherence peaks are smeared 
and the intra-gap spectra are broadened, with non-vanishing zero bias spectral weight, that we attribute to the 
metallic background. The curves are labelled by colored
triangles that refer to the corresponding temperatures in panel (a).}
\label{fig4}
\end{figure*}

Our present findings cast in a somewhat different perspective the analysis carried out in 
Ref.\,[\onlinecite{espci2}], where the appearance of more mobile carriers around $V_g=0$\,V was found. 
While on the one hand we fully confirm that result, on the other hand we find here that the magneto-transport 
data are well described assuming that these carriers do not appear uniformly in the whole system.
Rather high- and 
low-mobility carriers coexist in a high-mobility metallic phase, here identified with the phase (2), which 
is inhomogeneously distributed at the interface and spatially separated from the less metallic phase (1), where only 
one species of less mobile carriers is present. In this framework, the enhanced conductivity around $V_g=0$\,V and 
the changes of slope of the Hall resistance at high magnetic field occur because a finite fraction $w$ of phase 
(2), characterized by an overall higher carrier density and hosting high-mobility carriers, appears around this 
gate voltage [see Fig.\,\ref{fig3}(a)]. In agreement with the conclusions of Ref.\,[\onlinecite{espci2}], we also find 
that the more mobile carriers have a lower density than the low-mobility ones [Fig.\,\ref{fig3}(b)] but, 
at odds with the results obtained within a homogeneous description, in the 
present inhomogeneous scheme the mobilities of the two species stay rather constant over the whole range of gate 
voltages [see Fig.\,\ref{fig3}(c)].

\section{Percolative superconductivity} 
\label{sec3}
After assessing the occurrence of inhomogeneous multi-carrier magneto-transport at LXO/STO interfaces,
we discuss the superconducting transition that is driven by tuning the gate voltage $V_g$ (i.e., the
carrier density) above a threshold value. It has been shown that
the superconducting transition occurring in inhomogeneous systems is well described within the  
EMT.\cite{CGBC} EMT is a mean-field-like theory apt to describe a random resistor
network (RRN) that lacks spatial correlations. The EMT equations are obtained embedding one given random
resistance $R_i$ in an effective medium of constant resistance $R$. This latter is chosen in such a way
as to have the same current flowing through $R_i$ as in the RRN. The EMT resistance can be
shown\cite{CGBC} to be larger than the parallel and smaller than the series of the random resistances, 
the two limiting values being reached in infinite and one dimensions, respectively.

The resistance of the LTO/STO interface exhibits a marked suppression due to incipient 
superconductivity, that is accurately fitted\cite{rapid} assuming that the superconducting puddles occupy a 
fraction $w<1$ of the sample, and that each puddle is characterized by a random local critical temperature 
$T_c$. For the sake 
of definiteness, we adopt a Gaussian distribution of critical temperatures, $W(T_c)$, parametrized by its 
mean value $\overline T_c$ and its width $\gamma$. The remaining $1-w$ fraction of the sample is occupied by 
the metallic background. The resistance at temperature $T$ is found within EMT to be\cite{CGBC,BCCG,rapid}
\[
R(T)=R_\infty\left[(1-w)+w\,\mathrm{erf}\,\left(\frac{T-\overline T_c}{\gamma\sqrt{2}}\right)
\right],
\]
and results form the metallic background (first term inside the square brackets) and from not yet superconducting 
puddles (i.e., those puddles with $T_c<T$, second term inside the square brackets, erf being the error function); the
remaining puddles (i.e., those with $T_c>T$), have become superconducting and do not contribute to the
resistance. The high-temperature resistance $R_\infty$, $w$, $\overline T_c$, and $\gamma$ are used as fitting parameters. 
The global zero resistance state is reached at the percolative transition temperature $T_p\le \overline T_c$ such 
that
\begin{equation}
\mathrm{erf}\,\left(\frac{T_p-\overline T_c}{\gamma\sqrt{2}}\right)=\frac{w-1}{w}.
\label{weight}
\end{equation}
A solution for $T_p$ only exists if the superconducting fraction of the 2DEG can 
percolate in the two-dimensional system, i.e., for $w\ge\frac{1}{2}$. When $T_p<0$, or when it is not at 
all defined (for $w<\frac{1}{2}$), the resistance 
remains finite down to $T=0$, although the presence of a sizable (yet not percolating) 
superconducting fraction is mirrored by
a sizable suppression of $R(T)$, as sketched in Fig.\,\ref{fig4}(a).

\begin{figure}
\includegraphics[scale=0.3]{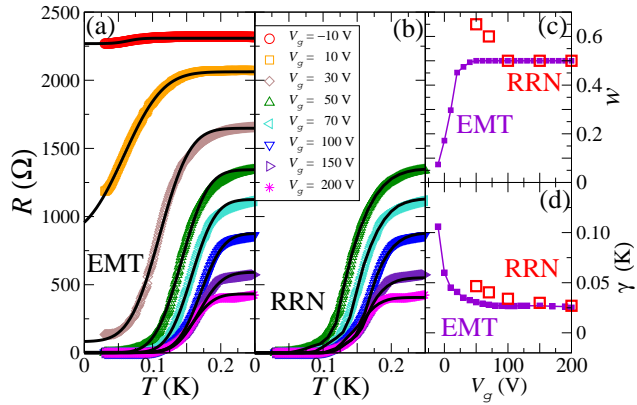}
\caption{(a) Measured sheet resistance in a LAO/STO sample, for $V_g=-10, 10, 30, 50, 70, 100, 150, 200$\,V. The 
solid lines are the EMT fits. (b) Same as in panel (a), with fewer experimental curves, fitted by RRN curves
(solid lines). (c) Weight of the superconducting fraction and (d) width of the $T_c$ distribution,
from EMT (line and small filled squares) and RRN (large open squares) fits.}
\label{fig5}
\end{figure}

The resistances measured in a LTO/STO sample as a function of $T$ for various values of the gate voltages 
$V_g$,\cite{espci2} are shown in Fig.\,\ref{fig5}(a) together with the fitting EMT curves (symbols and solid 
lines, respectively). Fig.\,\ref{fig5} is inspired to a similar figure of Ref.\,[\onlinecite{rapid}], 
with new elements included to make explicit contact with the overview contained in this piece of work.
The set of parameters $R_\infty,w,\overline T_c,\gamma$, i.e., the distribution of the puddles, changes 
with changing the carrier density by means of gating.
 
We are thus able to extract from the fits the mean intra-puddle critical temperature $\overline T_c$ [i.e., the
temperature at which $R(T)$ changes curvature within EMT], which will be analyzed in Sec.\,\ref{sec8}, 
and the overall superconducting 
fraction $w$, tracked by the solid line with filled squares in Fig.\,\ref{fig5}(c). We point out that EMT 
disregards spatial 
correlations, so that the presence of pronounced tails in the resistance, in the presence of a percolating 
superconducting cluster, force the overall superconducting fraction to be\cite{CGBC} 
$w\approx \frac{1}{2}$, as shown in Fig.\,\ref{fig5}(c). The width $\gamma$ of the 
Gaussian distribution of $T_c$ is plotted as a solid line with filled squares Fig.\,\ref{fig5}(d). It
increases as the fraction $w$ of the puddles goes to zero. This is rather natural because a reduction of the 
carrier density is expected to emphasize the effects of disorder, so that fluctuations of the local 
superconducting critical temperature should increase, leading to a broadening of the $T_c$ distribution.

A comment is now in order. Dealing with superconducting puddles embedded in a metallic background,
we expect the proximity effect\cite{proxeff} to play an important role in LXO/STO interfaces. 
Within EMT, this effect sure entails a temperature dependence of the superconducting fraction $w$. However,
when fitting the resistance curves $R(T)$, the introduction of a temperature dependent $w(T)$ is not 
viable, because it would made a good fit by definition. Nonetheless, the role of proximity effect can be
analyzed when discussing tunneling spectra. We shall come back to this point in Sec. \ref{sec6}.

\section{Superfluid density in inhomogeneous superconductors}
\label{sec4}
The inhomogeneous character of the 2DEG at the LXO/STO interfaces raises the question of the
description of the superfluid properties in a mixture of two phases.
The superfluid density $n_s$ was measured\cite{bert2012} by means of a SQUID in LAO/STO interfaces.
The measurements average over micrometric scales and are therefore not sensitive to submicrometric 
inhomogeneities. Nonetheless, the idea that the 2DEG at these interfaces is inhomogeneous, is still
supported by the evidence of variations of $n_s$ on the micrometric scale within a given 
sample. Encouraged by the marked similarity of the resistance curves in LAO/STO and in LTO/STO, we 
apply EMT also to describe the measured local $n_s$ as an average over an inhomogeneous state of 
submicrometric puddles. We point out that recent experiments in LTO/STO\cite{espci3} led to estimate the
typical size of the puddles to be $\approx 100$\,nm, thereby supporting this idea.

We proceeded extending EMT to small frequency $\omega$,\cite{rapid} and assigning to the metallic background 
a Drude-like complex conductivity $\sigma_M(\omega)=\Lambda\,(\Gamma+i\omega)^{-1}$ and to the superconducting 
puddles a purely reactive 
conductivity $\sigma_S(\omega)=\Lambda\,(i\omega)^{-1}$. We then defined the resistivity
$\rho_S(\omega)\equiv\sigma_S^{-1}(\omega)$ and 
$\rho_M(\omega)\equiv\sigma_M^{-1}(\omega)=\rho_0+\rho_S(\omega)$, with $\rho_0\equiv \Gamma/\Lambda$. At high $T$, 
the system is metallic and $\rho(\omega)=\rho_M(\omega)$. However, when the 
temperature is lowered, the static resistivity vanishes within each individual puddle as soon as $T$ equals 
the local $T_c$. 
Although the full expression of the complex resistivity can be found within EMT, aiming at describing the
static diamagnetic response, we give only the expression up to terms $\sim\omega$, i.e., 
$\rho(\omega)\approx\rho_0\,(w_M-w_S)\,\vartheta(w_M-w_S)+|w_M-w_S|^{-1}\rho_S(\omega)$,
where $\vartheta$ is the Heaviside function, $w_S$ is the fraction of puddles 
that have become superconducting (at a given temperature), and $w_M=1-w_S$ is the non superconducting fraction 
(resulting both from puddles that have not yet become superconducting and from the metallic background). 
Evidently, when $w_M>w_S$, the conductivity is Drude-like. However, below the percolation temperature 
$T_p$ (whenever defined), $w_M<w_S$, and the conductivity is purely reactive
$\sigma(\omega)=\Gamma\,(w_S-w_M)\,(i\omega)^{-1}$.
Therefore, employing also Eq.\,(\ref{weight}), we find that the superfluid density of the 
percolating two-dimensional network for $T\le T_p$ is
\begin{eqnarray}
n_s &\propto& w_S-w_M \nonumber\\
&=&w\left[ \mathrm{erf}\,\left(\frac{T_p-\overline T_c}{\gamma\sqrt{2}}\right)-
\mathrm{erf}\,\left(\frac{T-\overline T_c}{\gamma\sqrt{2}}\right)\right].
\label{stiffness}
\end{eqnarray}

\begin{figure}
\includegraphics[scale=0.3]{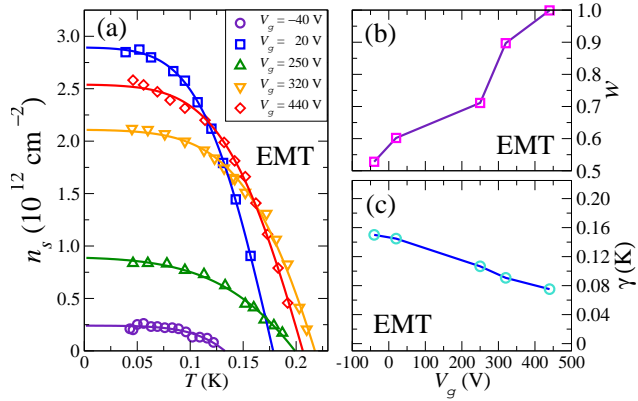}
\caption{(a) Superfluid density (symbols) as a function of $T$ and EMT
fits according to Eq.\,(\ref{stiffness}) (solid lines). Weight of the superconducting fraction (b) and width
of the $T_c$ distribution (c) extracted from the EMT fits, as a function of $V_g$. The lines are 
guides to the eye.}
\label{fig6}
\end{figure}

In Fig.\,\ref{fig6}(a), this average superfluid density and our EMT fits [Eq.\,(\ref{stiffness})] 
are reported, for different values of $V_g$. Fig.\,\ref{fig6} is inspired to a similar figure of 
Ref.\,[\onlinecite{rapid}], again suitably modified to make explicit contact with the present
discussion.
The behavior of $n_s(T)$, although qualitatively resembling 
the BCS prediction, may quantitatively differ from it. The slope at $T=T_p$, 
for instance, is controlled by the width $\gamma$ of the distribution $W(T_c)$. Thus, the deviations 
from standard BCS prediction measured in Ref. [\onlinecite{bert2012}], and attributed to a tendency to 
strong coupling, are here alternatively explained claiming that the 2DEG is in a weak coupling BCS regime 
(see Sec.\,\ref{sec8}), but the superfluid density is intrinsically inhomogeneous at the submicrometric scale 
and is averaged at the micron scale by the SQUID pick-up loop used in Ref. [\onlinecite{bert2012}]. 
We point out that our mean field approach is justified by the observation that the channels down to 500\,nm wide 
share properties similar to those of the larger samples.\cite{trisconeAPL} Our fits yield the fraction $w$ of 
volume occupied by the puddles, extracted from measurements of the superfluid density and reported in 
Fig.\,\ref{fig6}(b). It ranges from $\frac{1}{2}$ to 1, and is always larger than the fraction obtained from 
transport measurements. This is not surprising, because transport mainly probes the long-range connectivity of 
the percolating path, regardless of dead ends and disconnected superconducting regions,\cite{stauffer}, 
whereas diamagnetic screening measurements are sensitive to all sufficiently large superconducting loops, even 
when not connected to the backbone. In this case, the diamagnetic fraction can be large, while 
the long-distance connectivity is small, if many puddles or loops are disconnected (see Sec.\,\ref{sec5}). 
Fig.\,\ref{fig6}(c) reports the behavior of the width $\gamma$ of the distribution $W(T_c)$, inferred from the
superfluid density. We point out that, despite the fact that we are dealing with different materials and 
physical quantities, the behavior of $\gamma$ resembles that obtained from transport in LTO/STO 
[Fig.\,\ref{fig5}(d)] and is of comparable magnitude.

Of course, a proof of the above arguments requires a model accounting for space correlations, 
that includes both closed loops (relevant for diamagnetism) and connected paths (relevant for transport). 
While such a model has been found for transport (see Sec.\,\ref{sec5}), a similar model for 
the diamagnetic response is not yet available, so our discussion on this specific
aspect has a purely speculative character. An experimental test to our scenario could be provided by the 
observation of different diamagnetic 
responses in field-cooled and zero-field-cooled samples. In the first case one expects substantially smaller 
diamagnetic fraction, since a sizable flux would be trapped in the normal part encircled by the superconducting 
loops (see the shaded region in Fig.\,\ref{fig7} and the discussion in Sec.\,\ref{sec5}). 

The study of the dynamical response of an inhomogeneous superconductor at finite frequency, within EMT, 
is presently under investigation. A somewhat similar approach, applied to superconducting stripes in 
high-temperature superconductors, can be found in Ref. [\onlinecite{muniz}].

\section{Space correlations within the superconducting cluster} 
\label{sec5}
A possible explanation for the discrepancy between the superconducting fraction $w$ observed in transport and 
diamagnetism measurements may rest upon the filamentary structure of the superconducting cluster at oxide 
interfaces.\cite{rapid} So far, we have made use of the mean-field-like EMT, which completely neglects spatial 
correlations. To investigate the mechanisms determining the superconducting fraction observed in transport, 
$w\approx \frac{1}{2}$, we solved a RRN where the superconducting puddles form a spatially 
correlated cluster embedded in a metallic matrix. Preliminary results\cite{BCCG} indicated that a superconducting 
cluster which is dense at short distances and filamentary at larger distances is necessary in order to reproduce 
the observed tails of the resistance curves near percolation. 

\begin{figure}
\includegraphics[scale=0.4]{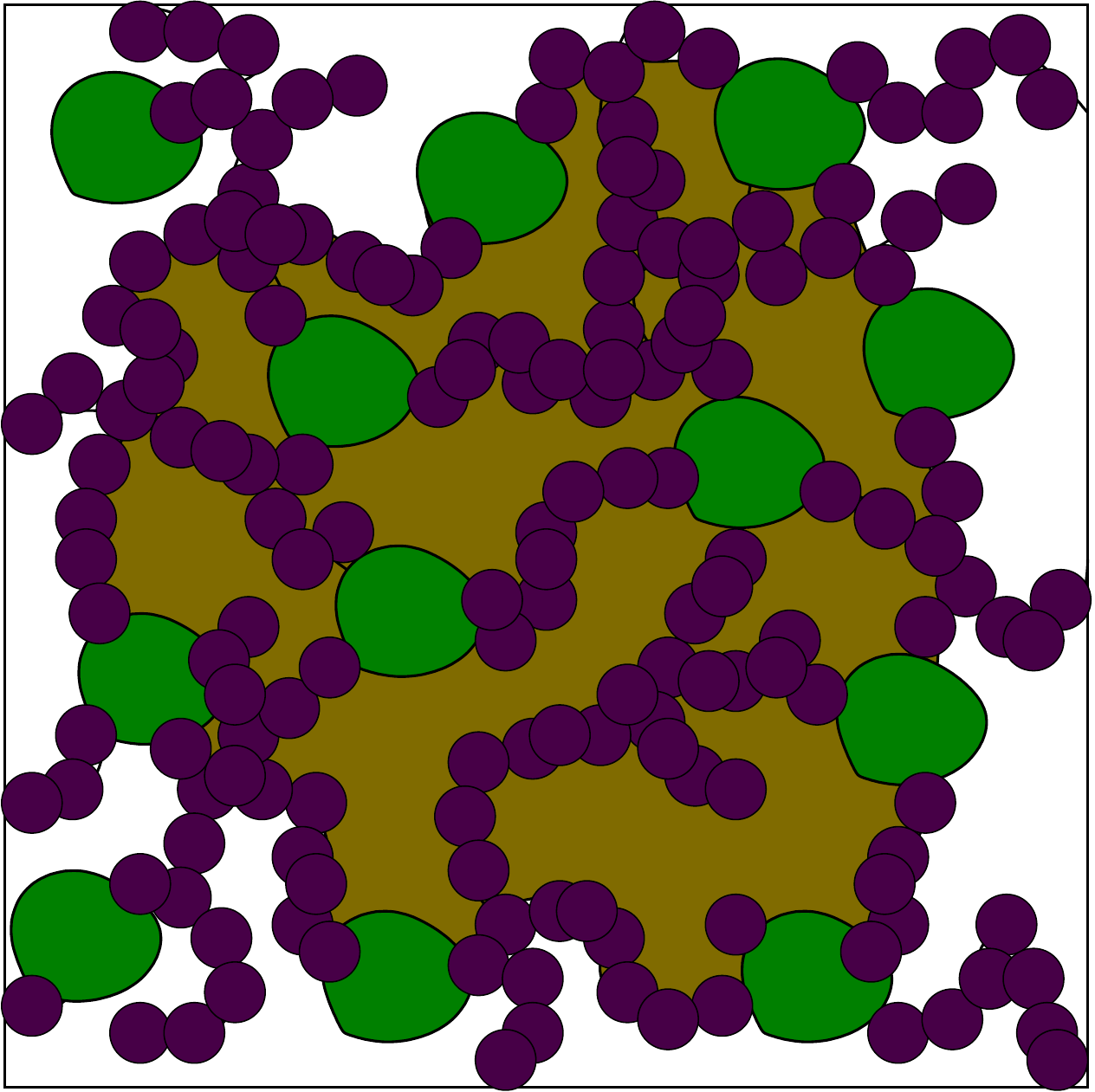}
\caption{Sketch of the RRN with superpuddles (green regions) connected by filaments of puddles (purple circles). 
The large shaded region, that does not contribute to superconducting transport, contributes instead to the 
diamagnetic response in the case of a zero-field-cooling measurement, since the superconducting currents confined
in the superpuddles and loop structures encircled by the puddles are apt to screen a magnetic field. In the case 
of a field-cooling measurement, one expects instead that the magnetic flux is frozen within the loop structures
encircled by the puddles.}
\label{fig7}
\end{figure}

To this purpose, we generated a fractal-like cluster with small long-scale connectivity, 
that percolates only when almost all bonds have become superconducting. We point out that the 
fractality of the clusters is an artifice to produce spatially correlated networks that are at the same time
dense at short distances and filamentary over long distances. For a Gaussian $W(T_c)$, the low temperature 
tail of the distribution must be necessarily reached, and a correspondingly pronounced tail in the  
resistance is obtained. Compact clusters fail to reproduce the tails in the resistance, in the absence of 
filamentary structures over long distances. Our systematic investigation showed that the presence of loosely connected
filaments is a necessary ingredient to reproduce the behavior of $R(T)$ near percolation.
On the other hand, a purely filamentary structure, no matter how dense at short distance, is too loose
and is not apt to describe the behavior of $R(T)$ at higher temperatures. In order to tune the density of the 
superconducting cluster without significantly changing the long-distance connectivity, we decorate the filaments 
with randomly distributed superpuddles, for simplicity assumed circular, their number and size being chosen to produce 
weights $w$ ranging from $0.3$ to $0.7$ (see the sketch in Fig.\,\ref{fig7}). Superpuddles may be produced by 
by large tails in the distribution of the puddle sizes, or by extrinsic pinning centers, promoting the nucleation
of much larger puddles. We systematically investigated the effect of the size and density of the superpuddles. A smaller 
fraction of larger superpuddles or a larger fraction of smaller superpuddles are essentially equivalent, as long as the 
superpuddles do not overlap to form percolating clusters.

In Fig.\,\ref{fig5}(b), we show our fits of the resistance obtained within our RRNs. Noticeably, even though 
the superconducting fraction $w$ is no longer forced to $\frac{1}{2}$, as it was instead within 
EMT, the resistance displays pronounced tails only if $0.50\lesssim w\lesssim 0.65$. The 
lower bound is imposed by the high slope at intermediate temperatures whereas the upper bound is due to the
pronounced tail near percolation.  In Figs.\,\ref{fig5}(c) and (d) we show the fraction $w$ occupied by the 
superconducting cluster and 
the width $\gamma$ of the Gaussian distribution $W(T_c)$ obtained within the RRN (open circles): $\gamma$ 
qualitatively resembles the width obtained within EMT, and increases upon lowering $V_g$. 
Thus, EMT and RRN models lead 
to similar results about the distribution of $T_c$ and its variation with $V_g$, provided the RRN is dense at 
short distances and filamentary over long distances.

\section{Tunneling spectra of inhomogeneous superconductors}
\label{sec6}
Another possibility of probing the inhomogeneous character of the supercondutcting state at LXO/STO
interfaces is provided by tunneling experiments. 
Recent metal-insulator-superconductor tunnel spectroscopy measurements on LAO/STO,\cite{tun} 
reveal the occurrence of a state with finite resistance, but superconducting-like density of states (DOS). 
The measurements are performed depositing a metallic Au electrode on top of the insulating 
LAO layer, and applying a bias voltage $V$ to drive a tunnel current $I$ between the electrode and the 2DEG. 
The electrode measures several hundreds $\mu$m across, thus being orders of magnitude larger than the 
nanoscale inhomogeneities.\cite{espci3} At the lowest measured temperature, $T=30$\,mK, the spectra reveal 
a gap in the DOS at the Fermi energy, over the entire range of explored gate voltages $V_g\in[-300,300]$\,V, 
accompanied by more or less broadened coherence peaks above the gap, pointing to superconducting coherence 
and pairing as 
the origin of DOS suppression [see Fig.\,\ref{fig4}(h)]. In the carrier depleted regime ($V_g\ll 0$), the 
suppression is found even in the absence of global superconductivity, 
again highlighting the inhomogeneous character of the state formed by superconducting puddles embedded in a 
metallic matrix. 
At very low carrier concentration, $V_g\approx -300$\,V, the coherence peaks have been completely
smeared, although a sizable gap is still present as a signature of (incoherent) pairing. We shall call
pseudo-gap state a state displaying at the same time a DOS suppression in the tunneling spectra, accompanied 
by more or less pronounced coherence peaks, and non-zero electrical resistance. At higher 
carrier concentration, $V_g=200$\,V, the 2DEG displays a superconducting gap and coherence peaks that decrease 
with increasing temperature and vanish around $300$\,mK (which agrees with the critical temperature of bulk 
STO reported in Ref. [\onlinecite{koonce}]). At $V_g\le 0$, the gap closes and the coherence peaks vanish, at 
temperatures much higher than the global $T_c$.

Since, as it was observed above, the tunneling spectra are taken over a spot measuring several hundreds 
$\mu$m across, we proposed\cite{noiPG} that the observed pseudo-gap results from an average over superconducting 
regions (the superconducting puddles, with a DOS described by standard BCS theory) and metallic regions 
(composed of the metallic background and of not yet superconducting puddles) with constant DOS $N_0$. 

The differential conductance is usually written as
\begin{equation}\label{dIdV}
\frac{\text{d}I}{\text{d}V}\bigl(V\bigr)=G_0-G_1\int_{-\infty}^{\infty}f'(E+eV)\,
N(E)\,\text{d}E,
\end{equation}
where $N(E)$ is the DOS, $f'(E)$ is the derivative of the Fermi distribution function, 
the positive constant $G_0$ customarily accounts for effects such as leakage currents, and $G_1$ is 
a dimensional constant. We recall that in our model superconducting pairing occurs within each puddle below a local 
critical temperature $T_c$, randomly distributed according to a probability distribution $W(T_c)$. The DOS of 
the 2DEG probed in tunneling spectra has three distinct contributions,
\begin{eqnarray}\nonumber
N(E)&=&(1-w)N_0+wN_0\int_{-\infty}^{T}\text{d}T_c\,W(T_c)\\ \label{rho}
&+&w\int_{T}^{\infty}\text{d}T_c\,W(T_c)\,N_{\Delta(T_c,T)}(E).
\end{eqnarray}
The first two terms correspond to the metallic background and to puddles where pairing has 
not taken place yet, respectively, and fully account for the zero bias background observed 
in the tunneling spectra, allowing us to take $G_0=0$ in Eq.\,(\ref{dIdV}). The third term corresponds 
to puddles that developed a finite pairing gap $\Delta$. The DOS within these puddles is taken as
\[
N_{\Delta}(E)=\biggl[(1-x)\,\frac{|E|}{\sqrt{E^2-\Delta^2}} + x\Biggr]\,N_0\,\vartheta\bigl(|E|-\Delta\bigr).
\]
The first term is the standard BCS expression, and describes 
coherent pairing occurring within a $(1-x)$ fraction of the gapped part. The second term describes puddles 
that, although having a gap, are too small to exhibit phase coherence. We define $w_{pair}$ as 
the total fraction of the system that can display pairing down to $T=0$, and the 
coherently and incoherently paired fractions $w_{coh}=(1-x)\,w_{pair}$ and $w_{inc}=x\,w_{pair}$. 
The latter term is necessary because the experimental spectra are gapped but display no coherence peaks
when $V_g\ll 0$.\cite{tun} The gap is assumed to take the BCS expression
\[
\Delta(T_c,T)=1.76\,T_c\,\tanh\biggl(\frac{\pi}{1.76}\sqrt{\frac{T_c-T}{T}}\,\biggr).
\]
The value of $N_0$ is readily determined by the high-bias part of the spectra, while $w$, $x$, and the
parameters of the distribution of critical temperatures, $\overline{T}_c$, $\gamma$, are 
used as fitting parameters.

At low temperature, accurate fits the spectra and of their evolution as a function of the
gate voltage $V_g$ were obtained.\cite{noiPG} Remarkably, the width of the distribution of critical temperatures,
$\gamma$, turned out to be very weakly dependent of $V_g$, indicating that the distribution of $T_c$
in the sample is essentially related to structural properties, like the local amount of disorder.

When fitting the temperature dependence of the spectra at fixed $V_g$,
the attempt to make use of the same temperature-independent set of parameters, although
capturing the main features of the spectra, yielded fits definitely less convincing than the fits at low 
temperature. However, releasing the severe constraint of temperature-independent parameters, very good fits are 
obtained by letting $w$ and $x$ vary, while the parameters of the distribution of critical temperatures, 
$\overline{T}_c$, $\gamma$ are kept fixed. The variation of $w$ and $x$ with the temperature is more clearly
understood when expressed through the quantities $w_{coh}$ and $w_{coh}$ defined above. It turns out that
the fraction of the sample occupied by the superconducting cluster increases with decreasing temperature, 
and saturates at low $T$, likely indicating that a sizable part of the metallic background is gradually 
driven superconducting by proximity effect.\cite{noiPG}

\section{Multiple quantum critical behavior in inhomogeneous superconductors}
\label{sec7}
\begin{figure}
\includegraphics[scale=0.35]{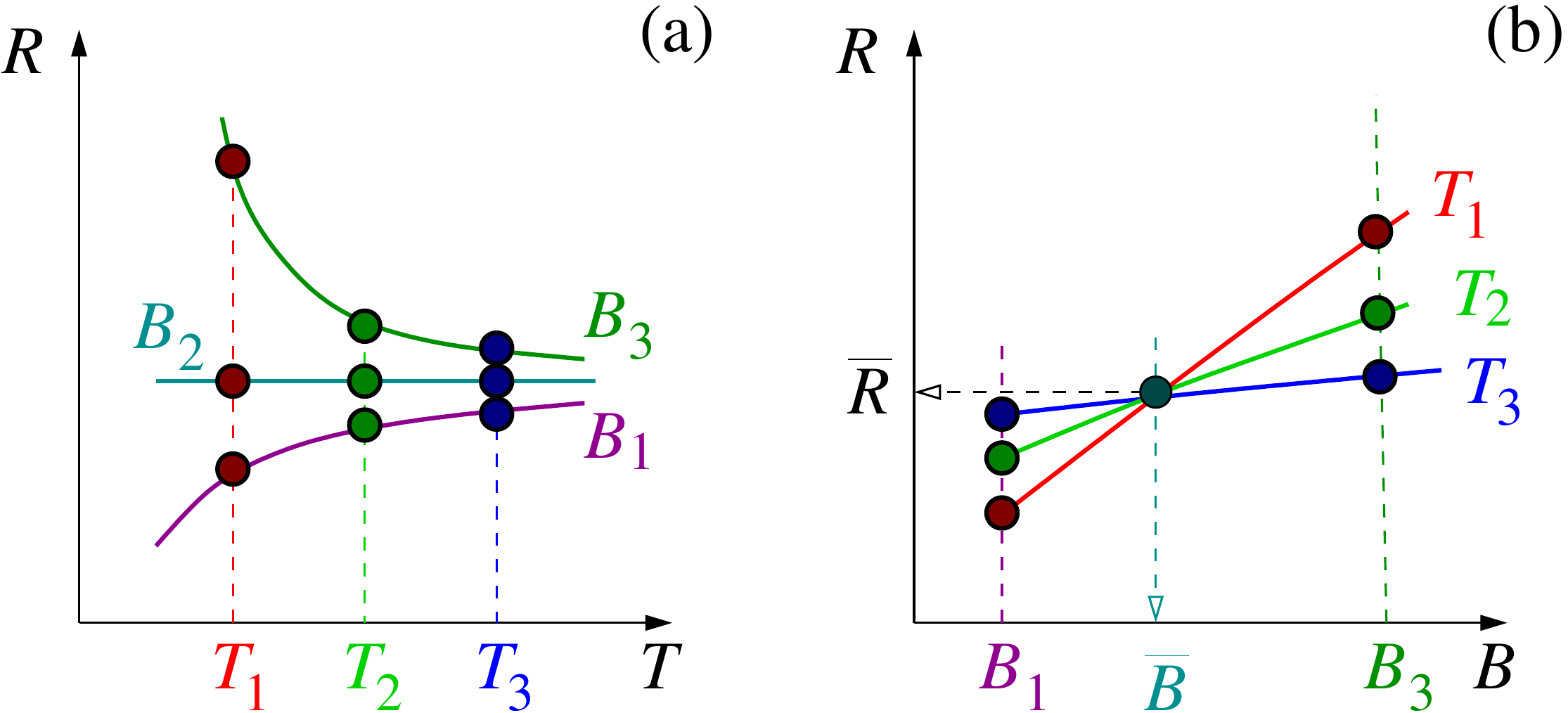}
\caption{Sketch of the procedure adopted to extract the quantum critical behavior when the superconducting phase
is suppressed by means of a magnetic field perpendicular to the interface. (a) When the resistance $R$ is plotted
as a function of temperature $T$, three behaviors are observed over a certain temperature interval, exemplified
by the curves corresponding to the three magnetic fields $B_1<B_2<B_3$: superconducting, critical, insulating.
(b) To better identify the critical value of the magnetic filed, $B_\times$ or $B_c$, isotherms are plotted
as a function of $B$, exemplified by the three curves corresponding to the temperatures $T_1<T_2<T_3$: the
crossing point of the isotherms corresponds to $\overline{B}$ ($B_\times$ or $B_c$), and 
the corresponding characteristic value $\overline R$ is obtained on the resistance axis.}
\label{fig8}
\end{figure}

The recent analysis of the quantum critical behavior that is observed when the superconducting phase is
suppressed by means of a magnetic field perpendicular to the interface\cite{espci3}
provided further support in favor of our scenario for inhomogeneous superconductivity
at oxide interfaces. When the resistance $R$ of LTO/STO, rescaled by a characteristic value
$\overline R$, is plotted as a function of the variable 
$(B-\overline{B})/T^{1/z\nu}$, where $\overline{B}$ is a characteristic magnetic field (see Fig.\,\ref{fig8}), 
$z$ is the dynamical exponent (converting a critical length scale in a time scale), and $\nu$ is the critical 
exponent of the correlation length as a function of temperature, two quantum 
critical scaling regimes are found, in different temperature
ranges. These two scaling regimes, separated by a crossover at intermediate temperatures, are in correspondence
with two different values of the characteristic magnetic field $\overline{B}=B_\times,B_c$. The smaller field, 
$B_\times$, is related to the scaling at higher $T$, with $z\nu\approx \frac{2}{3}$, whereas the (slightly) 
larger field, $B_c$, corresponds to the scaling at lower $T$, with $z\nu\approx \frac{3}{2}$. Noticeably, the 
characteristic fields $B_\times$ and $B_c$ coalesce at low carrier density (i.e., low gate voltage $V_g$). When
$V_g$ is increased, $B_\times$ saturates to a constant, whereas $B_c$ closely tracks the superconducting critical 
temperature $T_c$ (with a conversion factor of 1\,T corresponding to 1\,K).

A possible explanation for this multiple critical behavior relies on the assumption
that superconductivity within an isolated puddle would be suppressed by the smaller critical field $B_\times$. However, 
the puddles are coupled, being embedded in a common metallic background.\cite{sok} When inter-puddle coupling 
eventually intervenes, superconductivity is strengthened and survives up to the (slightly) larger critical field 
$B_c$. If we borrow the value of the dynamical exponent $z=1$, which is commonly adopted in
similar situations,\cite{zeta} the intra-puddle criticality is described by the  XY model\cite{xy} in 
the clean limit, where $\nu\approx \frac{2}{3}$. However, with lowering the temperature, the coherence length 
increases, eventually exceeding the puddle size, and inter-puddle superconductivity establishes in the inhomogeneous 
landscape of puddles embedded in the metallic background. In this case, the exponent $\nu$ must obey the Harris 
criterion\cite{harris} for disordered systems, and indeed we find $\nu\approx \frac{3}{2}>1$.

\section{Multi-carrier BCS model}
\label{sec8}
According to the discussion developed so far, the inhomogeneous character of the 2DEG at the LXO/STO oxide interfaces 
induces a distribution $W(T_c)$ of local critical temperatures, its mean value $\overline T_c$ depending on the overall 
carrier density (i.e., on $V_g$). This dependence is obtained fitting the resistance data in Fig.\,\ref{fig5}(a) 
within EMT, and sheds light on the intra-puddle pairing mechanism. As discussed in Sec.\,\ref{sec1}, 
detailed magneto-transport 
measurements\cite{espci2} highlighted the coexistence of a sizable amount of low-mobility carriers and a 
smaller amount of high-mobility carriers in LTO/STO. Here, as well as in 
LAO/STO,\cite{Rakhmilevitch} superconductivity seems definitely to develop
as soon as the high-mobility carriers appear.  

\begin{figure}
\includegraphics[scale=0.4]{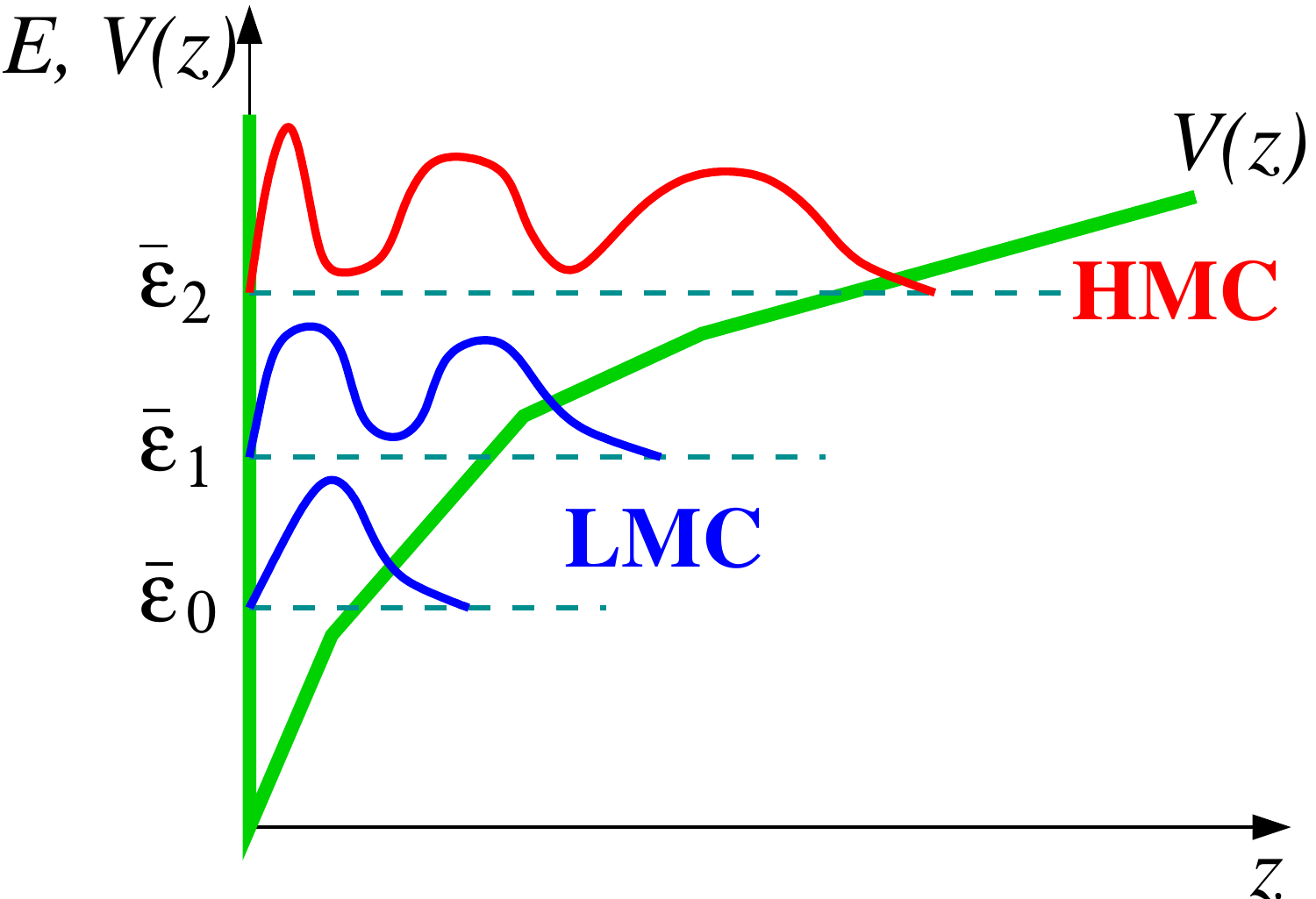}
\caption{Sketch of the quantized levels $\bar\varepsilon_\ell$ in the potential well $V(z)$ confining the 
2DEG at the interface ($z$ is the coordinate perpendicular to the interface), and of the corresponding carrier density
distributions. The lower-lying levels, e.g., at energies $E=\bar\varepsilon_0,\bar\varepsilon_1$, accommodate the 
low-mobility carriers (LMC), whereas the topmost level, e.g., at $E=\bar\varepsilon_2$, hosts the high-mobility 
carriers (HMC).}
\label{fig9}
\end{figure}

Accordingly, we proposed\cite{rapid} that superconducting pairing
within the 2DEG formed at the oxide interface may be described by a 
multi-band\cite{salluzzo,ilani,delugas,held,bianconi} BCS-like Hamiltonian
\begin{eqnarray}
\mathcal H_{BCS} &=&\sum_{\mathbf{k},\ell}
\xi_{\mathbf{k},\ell}\left(
a^\dagger_{\mathbf{k},\ell,\uparrow}a^{\phantom\dagger}_{\mathbf{k},\ell,\uparrow}
+a^\dagger_{\mathbf{k},\ell,\downarrow}a^{\phantom\dagger}_{\mathbf{k},\ell,\downarrow}
\right)
\nonumber\\
&+&{\sum_{\mathbf{k},\ell\atop \mathbf{k}',\ell'}}^\prime\, \frac{g_{\ell\ell'}}{N}\,
a^\dagger_{\mathbf{k},\ell,\uparrow}a^\dagger_{-\mathbf{k},\ell,\downarrow}
a^{\phantom\dagger}_{-\mathbf{k}',\ell',\downarrow}
a^{\phantom\dagger}_{\mathbf{k}',\ell',\uparrow}\label{eqbcs}
\end{eqnarray}
where $a^\dagger_{\mathbf{k},\ell,\sigma}
(a^{\phantom\dagger}_{\mathbf{k},\ell,\sigma})$ creates (annihilates) an electron
with two-dimensional wave vector $\mathbf{k}=(k_x,k_y)$, parallel to the plane of the
interface, and spin projection $\sigma=\uparrow,\downarrow$, belonging to the $\ell$-th sub-band, 
with dispersion law
\[
\xi_{\mathbf{k},\ell}=\bar\varepsilon_\ell+\frac{\hbar^2k^2_x}{2 m_{\ell,x}}
+\frac{\hbar^2k^2_y}{2 m_{\ell,y}}-\mu, 
\]
where $m_{\ell,x(y)}$ are the (possibly anisotropic) effective masses of 
the charge carriers and $\mu$ is the chemical potential, $g_{\ell\ell'}$ are the intraband (for $\ell=\ell'$) or
interband (for $\ell\neq\ell'$) pairing amplitudes and $N$ is the number of $\mathbf{k}$ points
within the first Brillouin zone. The sub-bands 
can originate, e.g., from the multiband structure of STO\cite{CPG,lungo} and/or from the quantum confinement within 
the self-consistent potential well at the interface.\cite{espci2,niccolo} In our schematic description,
we represent the whole set of low-lying bands with one 
sub-band ($\ell=1$) accommodating the non superconducting low-mobility carriers, while the high-mobility carriers 
in the sub-band $\ell=2$ are paired and give rise to a finite $T_c$ (see Fig.\,\ref{fig9},
where a sub-band structure originating from the quantum confinement is sketched, readapted 
from a similar sketch in Ref.\,[\onlinecite{rapid}]).

Thus, according to our interpretation, the superconducting puddles are regions where the 
$\ell=2$ sub-band is locally filled, whereas the (weakly localizing) metallic background corresponds to 
regions where the $\ell=2$ sub-band is empty. The phenomenology of the superconducting phase at oxide interfaces 
is reproduced assuming that the pairing amplitudes are such that
$g_{11}\ll (g_{12},g_{21})\ll g_{22}$. This condition is also consistent with the analysis of a two-band model 
in Ref. [\onlinecite{balatsky}]. For simplicity, to reduce to a minimum the number of free parameters, in the 
following we assume that $g_{11}=g_{12}=g_{21}=0$. According to the standard BCS approach, the pairing amplitudes 
are only effective in a window $|\xi_{\mathbf{k},\ell}|,|\xi_{\mathbf{k}',\ell'}|\le \hbar\omega_0$, where 
$\omega_0$ is a characteristic cut-off frequency. 
The prime superscript attached to the last sum in Eq. (\ref{eqbcs}) implies this restriction.
We assume that the bottoms of the two sub-bands are well separated,
$\bar\varepsilon_2-\bar\varepsilon_1\gg \hbar\omega_0$, and take henceforth $\bar\varepsilon_2=0$.

In principle, when deducing a BCS-like Hamiltonian, one should take care of vertex corrections
(which are expected to be relevant when Migdal's condition is violated). However, the task of deducing
a BCS-like hamiltonian in a multi-band model, where electrons with large and small Fermi energy coexist,
is overwhelmingly difficult. We rather take Eq. (\ref{eqbcs}) as a phenomenological low-energy effective 
Hamiltonian, imagining that the various high-energy effects have been accounted for by a suitable dressing 
of the bare physical parameters. Thus, the identification of $\omega_0$ with the characteristic frequency 
of the pairing mediator is expected to hold only indicatively.

\begin{figure}
\includegraphics[scale=0.25]{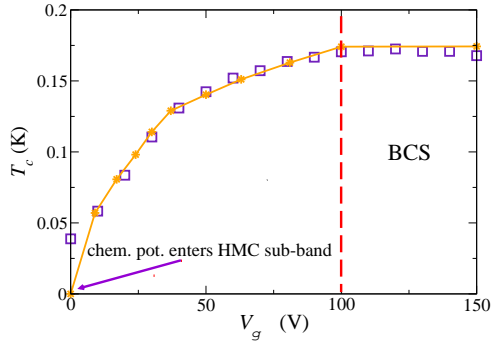}
\caption{Average critical temperature extracted from the EMT fit of the sheet resistance data 
(empty squares), and critical temperature of our BCS-like model (solid line with diamonds), as a function of $V_g$.
$T_c$ starts to rise when the chemical potential enters the HMC sub-band ($\mu=\bar\varepsilon_2$) and saturates to 
the standard BCS value when $\mu=\bar\varepsilon_2+\hbar\omega_0$. The two regimes are separated
by the dashed vertical line.}
\label{fig10}
\end{figure}

For $\mu\le 0$ the sub-band hosting the high-mobility carriers is empty and $T_c=0$. For $0<\mu<\hbar\omega_0$, 
pairing occurs at the critical temperature
\[
T_c\approx 1.14\,(\hbar\omega_0\mu)^{\frac{1}{2}}\,{\mathrm e}^{-1/\lambda},
\]
where $\lambda\equiv g_{22} N_{0}^\mathrm{HMC}$ is the dimensionless superconducting coupling and $N_{0}^\mathrm{HMC}$ 
is the DOS of the sub-band filled by the high-mobility carriers. For $\mu\ge\hbar\omega_0$, $T_c$ saturates to 
the standard BCS value
\[
T_c^{BCS}\approx 1.14\,\hbar\omega_0\,{\mathrm e}^{-1/\lambda}.
\]
The previous results can be cast into a single expression 
\begin{equation}
T_c(\mu)=T_c^{BCS}\vartheta(\mu)\min\left(\sqrt{\frac{\mu}{\hbar\omega_0}},1\right).
\label{tcmu}
\end{equation}
Remarkably, the amplitude $\Delta\mu$ of the interval in correspondence of which $T_c$ is an 
increasing function of $\mu$ provides a direct measure of the cut-off energy scale, 
$\hbar\omega_0=\Delta\mu$, which, with the caveat recalled above, we interpret as a possibly 
crude estimate of the characteristic energy scale of the pairing mediator. 

The fit of the curve $\overline T_c(V_g)$, extracted from the experimental data 
in Fig.\,\ref{fig5}(a) within EMT, with the curve $T_c(\mu)$ in Eq.\,(\ref{tcmu}), exploiting 
the approximately linear relation between $V_g$ and $\mu$,\cite{espci2} yields 
the result illustrated in Fig.\,\ref{fig10}
(orange line with diamonds and empty squares, respectively). Fig.\,\ref{fig10}
is inspired to a similar figure of Ref.\,[\onlinecite{rapid}], redrawn here in a modified fashion, 
to make explicit contact with the overview contained in this piece of work. The various band masses 
are taken all equal to 0.7 electron masses, corresponding to a scenario where the sub-band structure 
originates from quantum confinement of the lowest t$_{2g}$ band of bulk STO near the interface. 
We then obtain the dimensionless coupling constant $\lambda\approx 0.125$ 
(therefore, consistent with our assumption that the system falls in the weak coupling regime) 
and a cut-off energy scale $\hbar\omega_0\approx 23$\,meV, which is compatible with a typical phonon 
energy in STO.\cite{koonce}

\section{Concluding remarks}
\label{sec9}
In conclusion, we first analyzed magneto-transport experiments in LTO/STO oxide interfaces\cite{espci2} within
a multi-carrier EMT model, confirming the occurrence of two kind of carriers, with high and low mobility,
within an inhomogeneous landscape. We proposed that the system consists of regions with higher carrier density,
were both carrier coexist, and regions with lower carrier density, where only the low-mobility carriers are present. 
We also confirm that the high-mobility carriers have a lower density than the low-mobility ones. The 
mobilities of the two species turn out to be almost independent of the gate voltage. Thus, the enhancement of 
conductivity observed around $V_g=0$\,V and the change in the slope of the Hall resistance at high magnetic field 
occur because a finite fraction $w$ of regions with higher carrier density, hosting high-mobility carriers, appears 
around this gate voltage.

We then described superconductivity in LAO/STO and LTO/STO within a 
scenario in which superconducting puddles (the regions with coexisting high and low-mobility carriers) 
are embedded in a metallic background (the regions with low-mobility carriers only), and form a percolating network. 
In this framework, the sheet resistance of LTO/STO interfaces is very well described by EMT or by a RRN for 
an inhomogeneous 2DEG with a substantial filamentary character. Fitting the experiments, we were able
to extract the random distribution of $T_c$ at at various $V_g$ (i.e., various carrier densities ). A 
similar approach was adopted to fit the micrometrically averaged superfluid stiffness\cite{bert2012}
and the pseudo-gap in tunneling spectra\cite{tun} of LAO/STO. In particular, we showed that our model accounts 
well for the metal-insulator-superconductor tunneling spectra measured in LAO/STO, and allows us
to draw the conclusion that the fraction of the sample occupied by the superconducting puddles increases 
when the temperature is reduced, and saturates at low temperature, likely indicating that also a sizable 
fraction of the metallic background gradually becomes superconducting by proximity effect.

Assuming an effective two-band model with superconductivity triggered by the presence of few 
high-mobility carriers, locally filling the highest-energy band, we account for the density dependence of 
the intra-puddle $T_c$ within a simple BCS weak coupling scheme. As an important by-product, we find 
that the range of variation in $V_g$ of the (average) intra-puddle $T_c$ is directly related (via 
the chemical potential $\mu$) to the cut-off energy scale $\hbar\omega_0$. Taking this value as
a crude estimate of the typical energy scale of the pairing mediator, we find that this is
compatible with phonon-mediated superconductivity. 

Although this was not the main focus of the present piece of work, few words are now in order to discuss 
the origin of electron inhomogeneity at LXO/STO interfaces. Extrinsic mechanisms,\cite{extrinsic}
like impurities and growth
defects are always accountable of rendering an interface inhomogeneous. However, experimental data
show that inhomogeneity is never reduced below a still sizable extent, even in the best samples. 
Furthermore, the observation of negative electron compressibility in a low-filling regime\cite{negcompr} 
suggests that intrinsic mechanisms (e.g., in the form of effective electron-electron attractions) 
are present, which may render the 2DEG formed at the LXO/STO interface inhomogeneous by phase separation, 
even in a perfectly clean and expectedly homogeneous system. These facts
led us to look for intrinsic mechanisms of inhomogeneity. A possible intrinsic mechanism 
for the formation of an inhomogeneous 2DEG was proposed,\cite{CPG,lungo} that relies on the strong 
density dependent RSOC inferred at 
these interfaces,\cite{cavigliaprl} and yields a non-rigid band structure, with the possibility
that the chemical potential of the carriers is a non-monotonic function of the carrier density
for reasonable values of the model parameters, giving rise to a thermodynamic instability.
The resulting RSOC-driven electronic phase separation provides a natural 
framework for the occurrence of regions with higher and lower carrier density (the superconducting puddles and
the metallic background, respectively), accounting for the inhomogeneous character of the
superconducting phase at oxide interfaces. Of course, whenever electronic phase separation comes about,
one should worry about the Coulomb energy cost to be paid for the segregation of charge carriers. 
Thanks to the large dielectric constant of STO, we showed\cite{lungo} that the size of the domains can grow
as large as several tens of nanometers. An alternative route to intrinsic inhomogeneity might be provided by the 
pronounced density dependence of the self-consistent potential well $V(z)$ confining the 2DEG at the interface 
(see Fig.\,\ref{fig9}). Calculations within a Schr\"odinger-Poisson self-consistent scheme\cite{niccolo} show 
indeed that also this mechanism is apt to yield a non-monotonic chemical potential, again resulting in
electronic phase separation.

\par\noindent
~\vspace{0.5cm}
\par\noindent
{\bf Acknowledgments}. S.C. and M.G. acknowledge financial support from 
``AWARDS Projects'' of the University of Rome Sapienza, n. C26H13KZS9. This work has been supported 
by the Region \^ Ile-de-France in the framework of CNano IdF and Sesame programs, 
and by the DGA PhD program.


\begin{thebibliography}{99}

\bibitem{reyren} N. Reyren, S. Thiel, A. D. Caviglia, L. Fitting Kourkoutis,
G. Hammerl, C. Richter, C. W. Schneider, T. Kopp, A.-S. Rüetschi,
D. Jaccard, M. Gabay, D. A. Muller, J.-M. Triscone, and J. Mannhart,
Science {\bf 317}, 1196 (2007).

\bibitem{triscone} A. Caviglia, S. Gariglio, N. Reyren, D. Jaccard, T. Schneider, M. Gabay, S. Thiel, 
G. Hammerl, J. Mannhart, and J.-M. Triscone, Nature (London) {\bf 456}, 624 (2008).   

\bibitem{espci1} J. Biscaras, N. Bergeal, A. Kushwaha, T. Wolf, A. Rastogi, 
R. C. Budhani, and J. Lesueur, Nat. Commun. {\bf 1}, 89 (2010).  

\bibitem{espci2} J. Biscaras, N. Bergeal, S. Hurand, C. Grosset\^ete, A.
Rastogi, R. C. Budhani, D. LeBoeuf, C. Proust, and J. Lesueur, Phys. Rev. Lett. {\bf 108}, 247004 (2012).

\bibitem{ariando} Ariando, X. Wang, G. Baskaran, Z. Q. Liu, J. Huijben, J. B. Yi, A. Annadi, 
A. Roy Barman, A. Rusydi, S. Dhar, Y. P. Feng, J. Ding, H. Hilgenkamp, 
and T. Venkatesan, Nat. Commun. {\bf 2}, 188 (2011).

\bibitem{luli} Lu Li, C. Richter, J. Mannhart, and R. C. Ashoori, Nat. Phys. {\bf 7}, 762 (2011).

\bibitem{bert} J. A. Bert, B. Kalisky, C.  Bell, M. Kim, Y. Hikita, H. Y. Hwang, and K. A. Moler, 
Nat. Phys. {\bf 7}, 767 (2011).

\bibitem{metha1} D. A. Dikin, M. Mehta, C. W. Bark, C. M. Folkman, C. B. Eom, and V. Chandrasekhar, 
Phys. Rev. Lett. {\bf 107}, 056802 (2011).

\bibitem{metha2} M. M. Mehta, D. A. Dikin, C. W. Bark, S. Ryu, C. M. Folkman, C. B. Eom, 
and V. Chandrasekhar, Nat. Commun. {\bf 3}, 955 (2012).

\bibitem{bert2012} J. A. Bert, K. C. Nowack, B. Kalisky, H. Noad, J. R. Kirtley, C. Bell, H. K. 
Sato, M. Hosoda, Y. Hikita, H. Y. Hwang, K. A. Moler,  Phys. Rev. B {\bf  86}, 060503(R) (2012).

\bibitem{cavigliaprl} A. D. Caviglia, M. Gabay, S. Gariglio, N. Reyren, C. Cancellieri, and J.-M. Triscone, 
Phys. Rev. Lett. {\bf 104}, 126803 (2010).

\bibitem{CPG} S. Caprara, F. Peronaci, and M. Grilli, Phys. Rev. Lett. {\bf 109}, 196401 (2012).

\bibitem{rsoc}Y. A. Bychkov and E. I. Rashba, J. Phys. C {\bf 17}, 6039 (1984).

\bibitem{bell} C. Bell, S. Harashima, Y. Kozuka, M. Kim, B. G. Kim, Y.
Hikita, and H. Y. Hwang, Phys. Rev. Lett. {\bf 103}, 226802 (2009).

\bibitem{CGBC} S. Caprara, M. Grilli, L. Benfatto, and C. Castellani,
Phys. Rev. B {\bf 84}, 014514 (2011).

\bibitem{BCCG} D. Bucheli, S. Caprara, C. Castellani, and M. Grilli, 
New J. Phys. {\bf 15},  023014 (2013).

\bibitem{ristic} 
Z. Ristic, R. Di Capua, G. M. De Luca, F. Chiarella, G. Ghiringhelli, J. C. Cezar, N. B. Brookes, C. Richter, 
J. Mannhart and M. Salluzzo, Europhys. Lett. {\bf 93}, 17004 (2011).

\bibitem{feng_bi} F. Bi, M. Huang, C. W. Bark, S. Ryu, S. Lee, 
C.-B. Eom, P. Irvin, and J. Levy, arXiv:1302:0204.

\bibitem{Moller} B. Kalisky, private communication and APS March Meeting 2013.

\bibitem{ilani2} S. Ilani, private communication.

\bibitem{espci3} J. Biscaras, N. Bergeal, S. Hurand, C. Feuillet-Palma, A. Rastogi, R. C. Budhani,
M. Grilli, S. Caprara, and J. Lesueur,  Nat. Mater. {\bf  12}, 542 (2013).

\bibitem{rapid}
S. Caprara, J. Biscaras, N. Bergeal, D. Bucheli, S. Hurand, C. Feuillet-Palma, A. Rastogi, R. C. Budhani, 
J. Lesueur, and M. Grilli, Phys. Rev. B {\bf 88}, 020504(R) (2013).

\bibitem{tun} C. Richter, H. Boschker, W. Dietsche, E. Fillis-Tsirakis, R. Jany, F. Loder, L. F. Kourkoutis, 
D. A. Muller, J. R. Kirtley, C. W. Schneider, and J. Mannhart, Nature (London) {\bf 502}, 528 (2013). 

\bibitem{arkhincheev} V. E. Arkhincheev, JETP {\bf  91}, 407 (2000).

\bibitem{proxeff} Proximity effects in inhomogeneous superconductors are discussed in detail, e.g., in
Yu. N. Ovchinnikov, S. A. Wolf, and V. Z. Kresin, Phys. Rev. B {\bf  63}, 064524 (2001); 
V. Z. Kresin, Yu. N. Ovchinnikov, and S. A. Wolf, Phys. Rep. {\bf 431}, 231 (2006).

\bibitem{trisconeAPL} D. Stornaiuolo, S. Gariglio, N. J. G. Couto, A. F\^ete, 
A. D. Caviglia, G. Seyfarth, D. Jaccard, A. F. Morpurgo, and J.-M. Triscone,
Appl. Phys. Lett. {\bf 101}, 222601 (2012).

\bibitem{stauffer} D. Stauffer and A. Aharony, Introduction to Percolation Theory,
2nd ed. (Taylor and Francis, London, 1994).

\bibitem{muniz} R. A. Muniz and I. Martin, Phys. Rev. Lett. {\bf 107}, 127001 (2011).

\bibitem{koonce} C. S. Koonce, M. L. Cohen, J. F. Schooley, W. R. Hosler, and E. R. Pfeiffer, Phys. Rev. 
{\bf 163}, 380 (1967).

\bibitem{noiPG} D. Bucheli, S. Caprara, and M. Grilli, arXiv:1405.4666.

\bibitem{sok} B. Spivak, P. Oreto, and S. Kivelson,
Phys. Rev. B {\bf 77}, 214523 (2008).

\bibitem{zeta} S. L. Sondhi, S. M. Girvin, J. P. Carini, and D. Shahar, Rev. Mod. Phys. {\bf 69}, 315 (1997);
M. P. A. Fisher,  Phys. Rev. Lett. {\bf 65}, 923 (1990).

\bibitem{xy} D. Jasnow and M. Wortis, Phys. Rev. {\bf 176}, 739 (1968);
K. Wilson and M. Fisher, Phys. Rev. Lett. {\bf 28}, 240 (1972);
Y. H. Li, and S. Teitel, Phys. Rev. B {\bf 40}, 9122 (1989);
J. Kisker and H. Rieger, Phys. Rev. B {\bf 55}, R11981 (1997).

\bibitem{harris} H. B. Harris, J. Phys. C {\bf 7}, 1671 (1974).

\bibitem{Rakhmilevitch}
D. Rakhmilevitch, I. Neder, M. Ben Shalom, A. Tsukernik, M. Karpovski, Y. Dagan, and A. Palevski,
arXiv:1301.1055v1.

\bibitem{salluzzo} M. Salluzzo, J. C. Cezar, N. B. Brookes, V. Bisogni, G. M. De Luca, C. Richter, S. Thiel, 
J. Mannhart, M. Huijben, A. Brinkman, G. Rijnders, and G. Ghiringhelli, Phys. Rev. Lett. {\bf 102}, 
166804 (2009).

\bibitem{ilani} A. Joshua, S. Pecker, J. Ruhman, E. Altman and S. Ilani, Nat. Commun. {\bf 3}, 
1129 (2012).

\bibitem{delugas} P. Delugas, A. Filippetti, V. Fiorentini, D. I. Bilc, D. Fontaine, and P. Ghosez, 
Phys. Rev. Lett. {\bf 106}, 166807 (2011).

\bibitem{held} Z. Zhong, A. T\"oth, and K. Held,  Phys. Rev. B {\bf  87}, 161102(R) (2013).

\bibitem{bianconi} D. Innocenti, N. Poccia, A. Ricci, A. Valletta, S. Caprara, A. Perali, 
and A. Bianconi, Phys. Rev. B {\bf 82}, 184528 (2010);
D. Innocenti, S. Caprara, N. Poccia, A. Ricci, A. Valletta, and A. Bianconi, Supercond. Sci. 
Technol. {\bf 24}, 015012 (2011).

\bibitem{lungo} D. Bucheli, M. Grilli, F. Peronaci, G. Seibold, and S. Caprara, Phys. Rev. B 
{\bf 89}, 195448 (2014).

\bibitem{niccolo} N. Scopigno, D. Bucheli, S. Caprara, and M. Grilli, in preparation.

\bibitem{balatsky} R. M. Fernandes, J. T. Haraldsen, P. W\"olfle, and A. V. Balatsky, Phys. Rev. B 
{\bf 87}, 014510 (2013). 

\bibitem{extrinsic} N. C. Bristowe, T. Fix, M. G. Blamire, P. B. Littlewood, and E. Artacho, Phys. Rev. Lett. 
{\bf 108}, 166802 (2012).

\bibitem{negcompr} Lu Li, C. Richter, S. Paetel, T. Kopp, J. Mannhart, and R. C. Ashoori, Science {\bf 332}, 
825 (2011).

\end{thebibliography}
\end{document}